\documentclass[conference,a4paper]{APSIPA2021}
\usepackage{multirow,color}
\usepackage{graphicx}
\usepackage{amsmath}
\usepackage[psamsfonts]{amssymb}
\usepackage{amsxtra}
\usepackage{threeparttable}
\usepackage{cite}
\usepackage{color}

\usepackage{geometry}
\begin{document}

\title{Speech Enhancement Using Self-Supervised Pre-Trained Model and Vector Quantization}

\author{%
\authorblockN{%
Xiao-Ying Zhao\authorrefmark{1}, Qiu-Shi Zhu\authorrefmark{2} and Jie Zhang\authorrefmark{2}
}
\authorblockA{%
\authorrefmark{1}University of Science and Technology of China (USTC), Hefei, Anhui, China \\}
\authorblockA{%
\authorrefmark{2}National Engineering Research Center for Speech
and Language Information Processing,\\
University of Science and Technology of China (USTC), Hefei, Anhui, China\\
E-mail: \{xyzhao1123, qszhu\}@mail.ustc.edu.cn, jzhang6@ustc.edu.cn}
}

\maketitle

\footnotetext{Jie Zhang is the corresponding author.}

\begin{abstract}
With the development of deep learning, neural network-based speech enhancement (SE) models have shown excellent performance. Meanwhile, it was shown that the development of self-supervised pre-trained models can be applied to various downstream tasks. In this paper, we will consider the application of the pre-trained model to the real-time SE problem.
Specifically, the encoder and bottleneck layer of the DEMUCS model are initialized  using the self-supervised pre-trained WavLM model,  the convolution in the encoder is replaced by causal convolution, and the transformer encoder in the bottleneck layer is based on causal attention mask.
In addition, as discretizing the noisy speech representations is more beneficial for denoising,  we utilize a quantization module to discretize the representation output from the bottleneck layer, which is then fed into the decoder to reconstruct the clean speech waveform.
Experimental results on the Valentini dataset and an internal dataset show that the pre-trained model based initialization can improve the SE performance and the discretization operation suppresses the noise component in the representations to some extent, which can further improve the performance.

\end{abstract}

\section{Introduction}
Speech enhancement (SE) aims to improve the perceptual speech  quality and intelligibility by removing the background noises contained in the noisy input signal, which is usually exploited  as a front-end pre-processing module in many applications, such as automatic speech recognition (ASR), speaker diarization, hearing aids~\cite{9053317, 9664313}.
With the development of deep learning,  neural network-based SE models~\cite{8707065,8701652,defossez2020real} can even outperform the traditional counterparts~\cite{1163209}, which can be divided into two categories: time-frequency domain~\cite{6932438,8462068,9103053,hu20g_interspeech} and time-domain methods~\cite{8707065,8701652,defossez2020real}.
In  the time-frequency domain, a masking matrix is usually first estimated via supervised  training and  then multiplied with the noisy spectrum to estimate the clean spectrum, which is finally transformed into the time domain to recover the clean speech.
For the time-domain SE models, a convolutional encoder-decoder or Unet~\cite{ronneberger2015u} framework can be utilized to predict the clean speech waveform directly from noisy speech waveforms, where long short-term memory (LSTM)~\cite{defossez2020real} or self-attention~\cite{9746169} is adopted to model the temporal information. It has been experimentally shown  that these methods can improve the speech quality/intelligibility to some extent.

In the speech community, self-supervised pre-training models have been developed rapidly recently, which are pre-trained using large amounts of unlabeled data and then transferred to downstream tasks, e.g., ASR, speaker recognition.
For example, contrastive predictive coding (CPC)~\cite{oord2018representation} was proposed to predict the future frames using a contrastive loss. 
Wav2vec2.0~\cite{NEURIPS2020_92d1e1eb} leverages contextual information to predict the information of the masked frames using a contrastive loss function.
HuBERT~\cite{9585401} performs offline clustering for the representation output from the middle layer of the model, which enables to directly predict the clustering labels at the masked positions.
On the basis of HuBERT, WavLM~\cite{chen2021wavlm} can achieve the state-of-the-art performance on SUPERB benchmark~\cite{yang2021superb} by using an utterance mixing training and more unlabeled data. 
It was shown that there are some other pre-trained models, e.g.,~\cite{liu2021tera,hsu2021robust,zhu2022noise} that are beneficial for downstream tasks in both clean and noisy scenes.

Further, self-supervised pre-trained models can also be used to improve the SE performance in literature.
In~\cite{9746303}, thirteen pre-trained models are applied to the SE task, which are taken as feature extractors to generates spectral masks to reconstruct the clean speech waveform. It was shown  that the high-level features extracted by self-supervised pre-trained models are also applicable to SE compared to traditional acoustic features.
In~\cite{hung2022boosting}, the SE performance is improved by using a combination of features extracted by a self-supervised model and traditional spectral features.
However, currently these methods can only applied to offline SE tasks and few of them considers the application of self-supervised pre-trained models to the real-time case.
In addition, as it was shown in~\cite{chung2021w2v,zhu2022joint} that representation clustering can improve the noise robustness of the ASR,  inspired by the clustering representation approach in HuBERT, it might be the case that discretizing the noisy speech representations is more beneficial for denoising as well as the reconstruction of clean speech waveforms.

\begin{figure}[!t]
	\centering
	\includegraphics[width=0.48\textwidth]{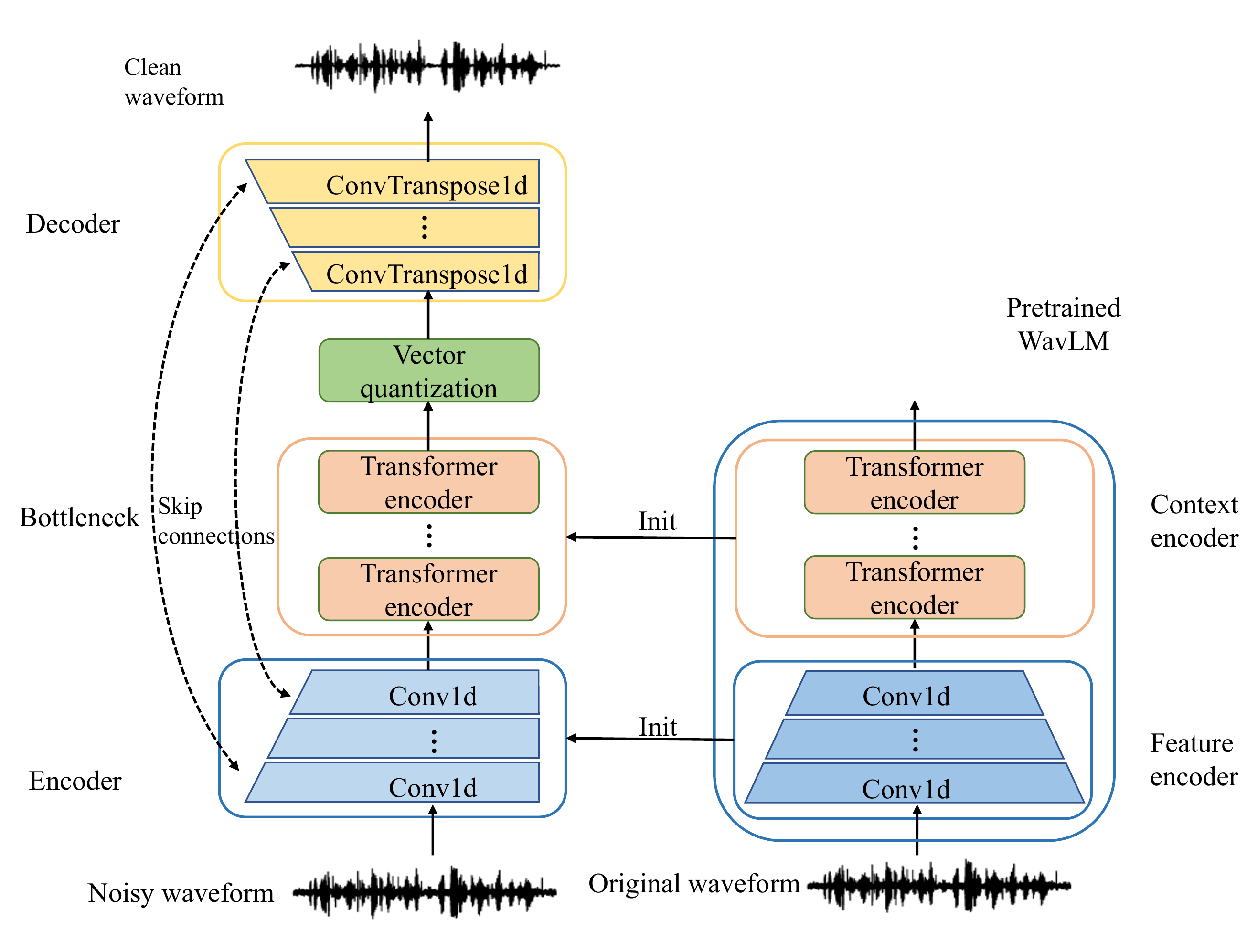}
	\caption{An illustration of the proposed speech enhancement model.}
	\label{fig:figure1}
\end{figure}

In this paper, we therefore consider the application of a self-supervised pre-trained model to the real-time SE task.
The overall configuration of the proposed model is shown in Fig.~\ref{fig:figure1}. The basis SE model adopts the Unet-based framework by initializing the encoder and bottleneck layers of the model using the pre-trained WavLM model and replacing the convolution of the encoder in the WavLM with the causal convolution. The causal attention mask is adopted for the Transformer encoder in the bottleneck layer.
In addition, We utilize a vector quantization (VQ) module to discretize the representation output from the bottleneck layer, which is then fed into the decoder for reconstructing clean speech waveforms.
Experimental results on the Valentini dataset~\cite{Valentini2017} and an internal dataset show that using the pre-trained model can improve the SE performance, and the proposed VQ module can further improve the performance.
Some enhanced speech samples and reference code can be found at: https://zxy0001.github.io.

\section{method}
\subsection{The self-supervised pre-trained WavLM model}
WavLM~\cite{chen2021wavlm} is a pre-training model based on HuBERT, which contains a convolutional encoder and a Transformer context encoder, where the convolutional encoder has several layers (each contains a time-domain one dimensional (1D) convolutional layer), a normalization layer, and a Gaussian error linear unit (GELU) activation layer.
In the Transformer context encoder, relative position information is introduced for the attention network by  gated relative position encoding to better model the local information.
In the pre-training stage, WavLM randomly transforms the input speech waveform, e.g., by mixing two waveforms or adding background noise. 
Afterwards, about 49\% of the speech signal is masked randomly in a sentence, and the discrete labels corresponding to the masked positions are predicted at the Transformer output, where the discrete labels are generated by discretizing the continuous signal via K-means clustering.
It was shown that WavLM can improves the performance of SUPERB benchmarks for various speech tasks with more unlabeled data and model parameters. 
For more details on WavLM, please refer to~\cite{chen2021wavlm}.

\subsection{The model architecture}
The proposed SE model is built based on the U-net structure, which contains an $encoder: X \mapsto Z$, a bottleneck $f: Z \mapsto C$, a $VQ: C \mapsto Q$ and a $decoder: Q \mapsto Y$, and the corresponding model structure is shown in Fig.~\ref{fig:figure1}.
The encoder has $D$ layers, and each contains a 1D causal convolution layer, a normalization layer and a GELU activation layer, where the convolution kernel size is $K$, the stride size is $S$ and the number of channels is $H$.
The decoder also has $D$ layers, and each contains a 1D causal transpose convolutional layer, a normalization layer and a GELU activation layer.
A skip connection is employed between the output of the $i$-th encoder and the input of the $i$-th decoder.
The bottleneck contains $N$ Transformer encoder layers, each of which contains a multi-head self-attention layer and a position-wise fully connected feed-forward layer. 
Skip connection and layer normalization are also utilized for each layer.

Specifically, given a noisy speech waveform $\boldsymbol{x}$, the feature $\boldsymbol{z} = encoder(\boldsymbol{x})$ is obtained by the encoder, which is then input to the bottleneck layer to obtain the contextual representation $\boldsymbol{c}=f(\boldsymbol{z})$.
Given the context representation $\boldsymbol{c}$, the discrete representation is calculated by the VQ module as $\boldsymbol{q}=VQ(\boldsymbol{c})$, which is finally exploited by  the decoder to reconstruct the enhanced speech waveform, i.e., $\boldsymbol{\hat{y}}=decoder(\boldsymbol{q})$.
As shown in Fig.~\ref{fig:figure1}, in order to ensure the causality of the proposed SE model (i.e., at time-frame $t$ only the information of previous frames can be used), we adopt a causal attention masking matrix to mask out the information after time $t$.
Since the convolutional receptive fields of the WavLM model and the SE model are different, we only utilize a few layers of the WavLM-base\footnote{https://github.com/microsoft/unilm/tree/master/wavlm} model to initialize the encoder and bottleneck of the SE model.

We obtain a finite set of representations by discretizing the bottleneck layer output via production quantization~\cite{5432202}, which is involved similarly to~\cite{NEURIPS2020_92d1e1eb}.
Given $G$ codebooks, each with $V$ learnable $d$-dimensional codewords, we first map the bottleneck representations $\boldsymbol{c}$ to $\boldsymbol{\rm l} \in \mathbb{R}^{G \times V}$ logits, and then select the discrete vectors by the Gumbel-softmax~\cite{jang2016categorical} operation in a differential way.
The probability of choosing the $v$-th codewords from the $g$-th codebook is given by
\begin{equation}
\overline{p}_{g,v}=\frac{\exp(\overline{l}_{g,v}+n_{v})/\tau}{\sum_{k=1}^{V}\exp(\overline{l}_{g,k}+n_k)/\tau},
\label{eq1}
\end{equation}
where $\tau$ is a non-negative temperature coefficient, $n_v=-\log(-\log(u))$ with $u$  uniformly distributed over [0, 1].
In the forward stage, the vector $i$ is selected as $i={\rm argmax}_{v}p_{g,v}$, and in the backward propagation stage the true gradient of the Gumbel-softmax output is utilized.
The quantization expects to utilize more codewords by maximizing the softmax distribution $\boldsymbol{\rm l}$, so we use the diversity loss function $L_d$, which is given by
\begin{equation}
	L_{d} = \frac{1}{GV}\sum_{g=1}^{G}\sum_{v=1}^{V}\overline{p}_{g,v}\log \overline{p}_{g,v}.
	\label{eq2}
\end{equation}

\subsection{The total loss function}
The total loss function $L_{\rm total}$ consists of an SE loss $L_{\rm se}$ and the diversity loss $L_d$, which reads
\begin{equation}
	L_{\rm total}=L_{\rm se} + \lambda L_{d},
	\label{eq3}
\end{equation}
where $\lambda$ is the balancing hyper-parameter. Note that the SE loss function $L_{\rm se}$ contains both the time-domain loss and the frequency-domain loss, which is similarly derived as in~\cite{defossez2020real,9746169}, 
where the time-domain component adopts $\ell_1$ loss and the frequency-domain component utilizes the multi-resolution STFT loss. As a result, $L_{\rm se}$  can be formulated as
\begin{equation}
	L_{\rm se} = \frac{1}{T} \left(\left\| \boldsymbol{y} - \boldsymbol{\hat{y}}  \right\|_1 + \sum_{i=1}^{M}L_{\rm stft}^{(i)}(\boldsymbol{y}, \boldsymbol{\hat{y}})\right),
	\label{eq4}
\end{equation}
where

\begin{align}
	L_{\rm stft}(\boldsymbol{y},\boldsymbol{\hat{y}}) &= L_{\rm sc}(\boldsymbol{y},\boldsymbol{\hat{y}}) + L_{\rm mag}(\boldsymbol{y},\boldsymbol{\hat{y}}),
	\label{eq5} \\
	L_{\rm sc}(\boldsymbol{y},\boldsymbol{\hat{y}}) &= \frac{\left\| | {\rm STFT}(\boldsymbol{y})| - | {\rm STFT}(\boldsymbol{\hat{y}})| \right\|_F}{\left\|  {\rm STFT}(\boldsymbol{y}) \right\|_F},
	\label{eq6}\\
	L_{\rm mag}(\boldsymbol{y},\boldsymbol{\hat{y}}) &= \frac{1}{T} \left\| \log |{\rm STFT}(\boldsymbol{y})|-\log |{\rm STFT}(\boldsymbol{\hat{y}})|  \right\|_1.
	\label{eq7}
\end{align}
where $M$ denotes the number of STFT loss functions and $\|\cdot\|$ the Frobenius norm. The multi-resolution loss $L_{\rm stft}^{(i)}$ utilizes the STFT loss with the number of FFT bins ranging from  \{512, 1024, 2048\}, hop size from \{50, 120, 240\} and window length from $\in$ \{240, 600, 1200\}, respectively.

\section{Experimental setup}
\subsection{Datasets and evaluation metrics}
The Valentini~\cite{Valentini2017} dataset contains 28.4 hours of clean and noisy speech pairs at a sampling rate of 48 kHz.
These data are collected from 84 speakers at 4 signal-to-noise ratios (SNRs) (i.e., 0, 5, 10 and 15 dB) in the training set and 4 SNRs (2.5, 7.5, 12.5 and 17.5 dB) in the test set.
We downsampled the raw speech waveforms to 16kHz and then applied the same remix augmentation and bandmask augmentation methods as in~\cite{defossez2020real}.
In addition, we also validate the proposed model using the internal challenging 100-hour noisy data.

In addition, in order to show the generalizability of the proposed model, we also evaluate the performance on a self-built dataset, which is more challenging than the Valentini dataset in principal.
The training set of the internal dataset involves 100 hours data from Librispeech clean and noisy speech~\cite{7178964}, where the noisy mixtures are generated by mixing the clean speech and the noise sources from the Freesound~\cite{font2013freesound} noise dataset at a SNR randomly chosen from [0, 10] dB.
The testing set has 15 hours speech consisting of 5000 utterances from 10 speakers recorded in the real traffic and in-car environments. The sampling frequency is 16 kHz.

The performance of the proposed method is evaluated using objective metrics, including 1) perceptual evaluation of speech quality (PESQ)~\cite{pesq}, 2) short-time objective intelligibility (STOI)~\cite{5713237}, 3) mean opinion score (MOS) prediction of distortion of speech signal (SIG)~\cite{4389058}, 4) MOS prediction of intrusiveness of background noise (BAK)~\cite{4389058}, 5) MOS prediction of overall quality (OVRL)~\cite{4389058}. 
Some enhanced speech samples and reference code can be downloaded from {https://zxy0001.github.io}.

\subsection{Model configuration}
In order to enable the initialization of the proposed SE model using the pre-trained WavLM, the entire SE model utilizes a 3-layer ($D$ = 3) 1D convolution for the encoder, a 2-layer ($N$ = 2) Transformer encoder for the bottleneck and a 3-layer 1D transposed convolution for the decoder.
Both the encoder and decoder have a dimension of 512 ($H$ = 512). The Transformer encoder in the bottleneck layer has a dimension of 768, the feedforward neural network has a dimension of 2048, and the self-attention module has 12 heads.
For the 1D convolution in encoder, the convolution kernel size and stride size are (10, 3, 3) and (5, 2, 2), respectively.
For the VQ module, we adopt $G$ = 1 codebook with $V$ = 320 learnable 128-dimensional vectors in each codebook, and $\lambda$ in (\ref{eq3}) is set to be 0.01 due to the range of the diversity loss.
We use the Adam optimizer to train the SE model for 1M iterations, where the batch size is 64 and the maximum learning rate is $2 \times 10^{-4}$. All models are trained on 4 Tesla-V100-32G GPUs.

\begin{table*}[]
     \caption{Performance comparison of objectively evaluation metrics on the Valentini test set.}
     \label{tab:table1}
     \centering
	\begin{tabular}{l|c|ccccc|c}
		\hline
		\textbf{Model}  & \textbf{Domain}    & \begin{tabular}[c]{@{}c@{}}\textbf{PESQ (WB)}\end{tabular} & \begin{tabular}[c]{@{}c@{}}\textbf{STOI (\%)}\end{tabular} & \begin{tabular}[c]{@{}c@{}}\textbf{pred.}\\ \textbf{CSIG}\end{tabular} & \begin{tabular}[c]{@{}c@{}}\textbf{pred.}\\ \textbf{CBAK}\end{tabular} & \begin{tabular}[c]{@{}c@{}}\textbf{pred.}\\ \textbf{COVL}\end{tabular} & \textbf{Causality} \\ \hline
		Noisy  & -    & 1.97                                                & 92.1                                                & 3.35                                                 & 2.44                                                 & 2.63                                                 & -      \\ \hline
		SEGAN~\cite{Pascual2017}  & waveform     & 2.16                                                & -                                                   & 3.48                                                 & 2.94                                                 & 2.80                                                 & No     \\
		Wave U-Net~\cite{macartney2018improved} & waveform  & 2.40                                                & -                                                   & 3.52                                                 & 3.24                                                 & 2.96                                                 & No     \\
		SEGAN-D~\cite{9201348} & waveform    & 2.39                                                & -                                                   & 3.46                                                 & 3.11                                                 & 3.50                                                 & No     \\
		MMSE-GAN~\cite{8462068}  & time-frequency  & 2.53                                                & 93.0                                                  & 3.80                                                 & 3.12                                                 & 3.14                                                 & No     \\
		MetricGAN~\cite{pmlr-v97-fu19b}  & waveform  & 2.86                                                & -                                                   & 3.99                                                 & 3.18                                                 & 3.42                                                 & No     \\
		DeepMMSE~\cite{9066933}  & waveform  & 2.95                                                & 94.0                                                  & 4.28                                                 & 3.46                                                 & 3.64                                                 & No     \\
		DEMUCS~\cite{defossez2020real}  & waveform    & 3.07                                                & 95.0                                                  & 4.31                                                 & 3.40                                                  & 3.63                                                 & No     \\ 
		CleanUNet~\cite{9746169} & waveform          & 3.09             & 95.8                     & 4.38          & 3.47   & 3.69 & No   \\
		\hline
		Ours (D=3, N=2) & waveform      & \textbf{3.13}                                                & \textbf{96.1}                                                & \textbf{4.46}                                                 & \textbf{3.56}                                                 & \textbf{3.82}                                                 & No     \\ \hline
		\hline
		Wiener &-     & 2.22                                                & 93.0                                                  & 3.23                                                 & 2.68                                                 & 2.67                                                 & Yes    \\
		DeepMMSE~\cite{9066933} & waveform   & 2.77                                                & 93.0                                                  & 4.14                                                 & 3.32                                                 & 3.46                                                 & Yes    \\
		DEMUCS~\cite{defossez2020real} & waveform     & 2.93                                                & 95.0                                                  & 4.22                                                 & 3.25                                                 & 3.52                                                 & Yes    \\
		CleanUNet~\cite{9746169} & waveform  & 2.91                                                & 95.6                                                & 4.34                                                 & 3.42                                                 & 3.65                                                 & Yes    \\ \hline
		Ours (D=3, N=2)   & waveform     & \textbf{3.02}                                                & \textbf{95.8}                                                & \textbf{4.40}                                                 & \textbf{3.49}                                                 & \textbf{3.72}                                                 & Yes    \\ \hline
	\end{tabular}
\end{table*}

\section{Experimental results}

\textbf{Comparison methods: } We first measure objective evaluation metrics on the noisy test set as a baseline. The time-domain SE comparison methods include SEGAN~\cite{Pascual2017}, SEGAN-D~\cite{9201348} and MetricGAN~\cite{pmlr-v97-fu19b}, which are based on the  generative adversarial networks (GAN). The time-frequency domain SE comparison approaches include MMSE-GAN~\cite{8462068}, which is also relied on the GAN model, and DeepMMSE~\cite{9066933}, which estimates the noise spectral density using the minimum mean square error (MMSE) criterion.  Wave U-Net~\cite{macartney2018improved}, DEMUCS~\cite{defossez2020real} and CleanUNet~\cite{9746169} are also compared, which utilize the Unet network for the time-domain SE.

Table~\ref{tab:table1} shows the experimental results of the comparison methods on the Valentini dataset. It is clear that the proposed model achieves a PESQ of 3.13 and an STOI of 96.1\%, respectively using the pre-trained WavLM for initialization and the extra quantization module, which outperforms DEMUCS and CleanUNet.
For the real-time SE case, our model achieves a PESQ of 3.02 and an STOI of 95.8\%, respectively, which are also better than the best off-the-shelf models, e.g., DEMUCS and CleanUNet.
As the  average input  SNR of the Valentini dataset is moderately high, in order to show the generality of the proposed method we employed a more challenging internal dataset for testing, and the obtained results are shown in Table~\ref{tab:table2}. As CleanUNet and DEMUCS obtain the best performance among all chosen comparison methods, we will only compare them with the proposed method in the sequel.
We can see that the performance of the proposed method with random initialization is comparable to that of the CleanUNet model, since both employ self-attention modules as bottleneck and the amounts of model parameters are roughly equal.
In case WavLM is adopted to initialize our model, the performance is clearly improved compared to the random initialization. Note that using the VQ module in addition to the WavLM-based initialization does not always mean a positive performance gain, which should be dependent on the number of learnable codewords. When the context representation is quantized with sufficient codewords, the proposed VQ module can further improve the SE performance in terms of all metrics (e.g., Init V = 480 vs. Init). This is due to the fact that in case the number of codewords is small, the quantization noise will heavily affect the context representation as well as the diversity loss function and a certain loss in speech context information becomes inevitable. In general, the less the codewords, the more the information loss and the higher the quantization noise variance.

\begin{table}[]
	\caption{Performance comparison of objective evaluation metrics for real-time speech enhancement on the internal test set.}
	\label{tab:table2}
	\centering
	\begin{tabular}{l|ccccc}
		\hline
		\textbf{Model}              & \begin{tabular}[c]{@{}c@{}}\textbf{PESQ}\\ \textbf{(WB)}\end{tabular} & \begin{tabular}[c]{@{}c@{}}\textbf{STOI}\\ \textbf{(\%)}\end{tabular} & \begin{tabular}[c]{@{}c@{}}\textbf{pred.}\\ \textbf{CSIG}\end{tabular} & \begin{tabular}[c]{@{}c@{}}\textbf{pred.}\\ \textbf{CBAK}\end{tabular} & \begin{tabular}[c]{@{}c@{}}\textbf{pred.}\\ \textbf{COVL}\end{tabular} \\ \hline
		Noisy              & 1.32                                                & 80.8                                                & 2.54                                                 & 1.87                                                 & 1.86                                                 \\ \hline
		DEMUCS~\cite{defossez2020real}             & 1.89                                                & 86.2                                                & 3.25                                                 & 2.37                                                 & 2.50                                                 \\
		CleanUNet~\cite{9746169}          & 2.17                                                & 87.1                                                & 3.61                                                 & 2.54                                                 & 2.62                                                 \\ \hline
		Ours (random init) & 2.04                                                & 86.5                                                & 3.53                                                 & 2.49                                                 & 2.55                                                 \\
		Ours (Init)        & 2.29                                                & 88.1                                                & 3.70                                                 & 2.61                                                 & 2.72                                                 \\
		Ours (Init, V=160) & 2.20                                                & 87.3                                                & 3.65                                                 & 2.59                                                 & 2.65                                                 \\
		Ours (Init, V=320) & 2.35                                                & 88.6                                                & 3.79                                                 & 2.66                                                 & 2.78                                                 \\
		Ours (Init, V=480) & \textbf{2.37}                                                & \textbf{88.9}                                                & \textbf{3.80}                                                 & \textbf{2.67}                                                 & \textbf{2.81}                                                 \\ \hline
	\end{tabular}
\end{table}

\begin{figure}[!t]
	\centering
	\includegraphics[width=0.45\textwidth]{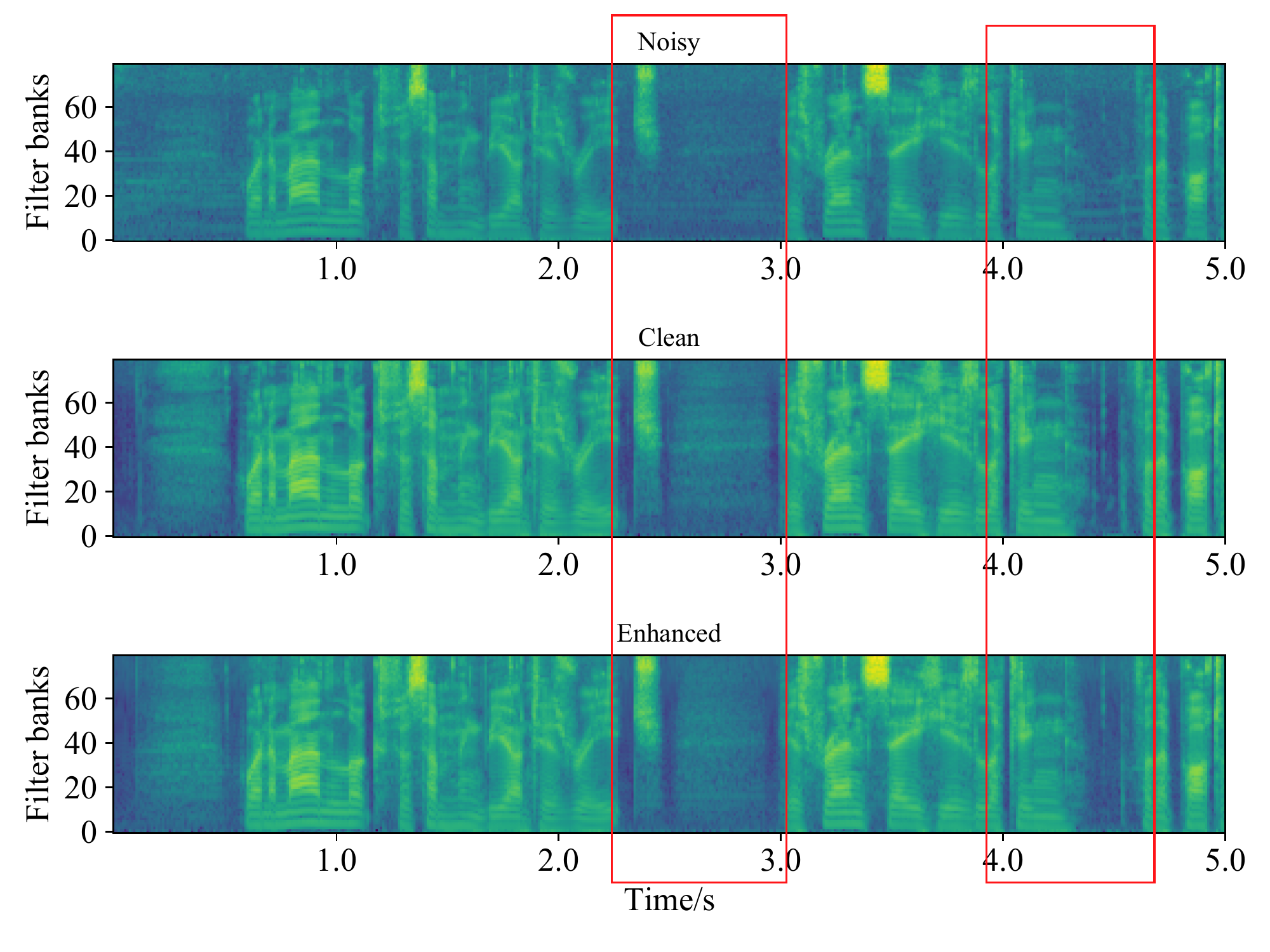}
	\caption{A filterbank feature of an example speech.}
	\label{fig:figure2}
	\vspace{-0.3cm}
\end{figure}

This can also be seen from different choices of $V$ in Table~\ref{tab:table2}, as the SE performance of the proposed model decreases in case the number of learnable vectors decreases.
Finally, we visualize the filterbank features of speech signals in Fig.~\ref{fig:figure2}, from which it can be clearly seen that SE can remove the noise component from the noisy speech to some extent.
In addition, for the causal models we evaluate the real-time factor (RTF), which is defined as the required processing time of unit-second speech. The RTF is computed on a 12-core Intel E5-2680 v3 CPU. We find that DEMUCS, CleanUet and the proposed method obtain a comparable RTF of 0.66, which tells that the proposed model does not increase the time complexity.

\section{Conclusions}

In this paper, we investigated the effect of using pre-trained models for initialization and vector quantization for context representations on the real-time SE problem.
It was shown that the pre-trained WavLM based initialization for the encoder and bottleneck layers of the SE model can always lead to a performance improvement compared to the random initialization, while the utilization of the VQ module should be careful. Only when the size of representation codebook is large enough (as the quatization noise does not dominate), it can further improve the SE performance on the basis of the WavLM-based initialization. 
In principal, applying the quantization module to discretize the representation of the bottleneck layer output functions as a preliminary noise reduction processor on the speech representation.

\section*{Acknowledgment}
This work was supported by the National Natural Science Foundation of China (62101523), Hefei Municipal Natural Science Foundation (2022012) and Fundamental Research Funds for the Central Universities.

\bibliographystyle{IEEEbib}
\bibliography{refs}

\end{document}